\documentclass[pdflatex,sn-mathphys-num]{sn-jnl}

\usepackage{graphicx}
\usepackage{multirow}
\usepackage{amsmath, amssymb, amsfonts}
\usepackage{amsthm}
\usepackage{mathrsfs}
\usepackage[title]{appendix}
\usepackage{xcolor}
\usepackage{textcomp}
\usepackage{manyfoot}
\usepackage{booktabs}
\usepackage{listings}
\usepackage{tabularx}
\usepackage{array}
\usepackage{caption}
\usepackage{subcaption}

\usepackage{algorithm}
\usepackage{algorithmic}
\usepackage{tikz}
\usetikzlibrary{shapes, arrows, positioning}
\theoremstyle{thmstyleone}

\theoremstyle{thmstyletwo}

\theoremstyle{thmstylethree}

\begin{document}

\title[Article Title]{ Agentic AI for Education: A Unified Multi-Agent Framework for Personalized Learning and Institutional Intelligence }


\author*[1]{\fnm{Arya Mary } \sur{K J}}\email{aryamary@mec.ac.in}
\author*[1]{\fnm{Deepthy} \sur{K Bhaskar}}\email{dmdl22jan005@mec.ac.in}
\author[1]{\fnm{Sinu } \sur{T S}}\email{sinuts@mec.ac.in}

\author[1]{\fnm{Binu} \sur{V P}}\email{binuvp@mec.ac.in}

\affil*[1]{\orgdiv{Department of Computer Engg}, \orgname{Govt. Model Engineering College,APJ Abdul Kalam Technological University}, \orgaddress \city{Kerala},  \country{India}}


\abstract{Agentic Artificial Intelligence (AI) represents a paradigm shift from reactive systems to proactive, autonomous decision making frameworks. Existing AI-based educational systems remain fragmented and lack multi-level integration across stakeholders. This paper proposes the Agentic Unified Student Support System (AUSS), a novel multi-agent architecture integrating student-level personalization, educator-level automation, and institutional-level intelligence. The framework leverages Large Language Models (LLMs), reinforcement learning, predictive analytics, and rule-based reasoning. Experimental results demonstrate improvements in recommendation accuracy (92.4\%), grading efficiency (94.1\%), and dropout prediction (F1-score: 89.5\%). The proposed system enables scalable, adaptive, and intelligent educational ecosystems.

}

\keywords{Agentic AI, Multi-Agent Systems, Adaptive Learning, Reinforcement Learning, Educational Technology, Learning Analytics
}



\maketitle

\section{Introduction}

The rapid advancement of Artificial Intelligence has significantly transformed the educational landscape, enabling the development of intelligent tutoring systems, automated grading mechanisms, and data-driven learning analytics. These technologies have improved accessibility and scalability, allowing institutions to deliver education more efficiently while supporting diverse learning needs. Recent studies have highlighted the growing role of AI in enhancing adaptive learning environments and improving student engagement through data-driven personalization~\cite{b1,b2}.

Despite these advancements, most AI-driven educational systems continue to operate in a predominantly reactive manner, responding only when prompted by users rather than proactively guiding the learning process. Such limitations restrict their ability to provide timely interventions and personalized support. Research has shown that traditional AI models often lack contextual awareness and adaptability, which are essential for addressing the dynamic nature of learning environments~\cite{b3,b4}.

In real-world educational settings, learning is influenced by multiple interconnected factors, including student behavior, instructional strategies, and institutional policies. Conventional systems struggle to integrate these dimensions effectively, resulting in delayed feedback, increased educator workload, and inefficient institutional decision-making. Recent works emphasize the need for intelligent systems capable of continuous monitoring and predictive analysis to improve educational outcomes~\cite{b5,b6}.

Agentic Artificial Intelligence has emerged as a promising paradigm that addresses these limitations by introducing autonomy, goal-driven reasoning, and continuous adaptation. Unlike conventional AI models, agentic systems possess the ability to independently plan tasks, execute decisions, evaluate outcomes, and refine their behavior over time. This shift toward autonomy enables proactive interventions and intelligent decision-making across complex environments. Studies in this domain have demonstrated the potential of agentic systems to enhance scalability, adaptability, and system-level intelligence~\cite{b7,b8}.

In this context, the integration of agentic AI into education presents a unique opportunity to move beyond fragmented and isolated solutions toward a unified and collaborative framework. This paper proposes AUSS, a multi-agent architecture designed to seamlessly integrate student-level personalization, educator-level automation, and institutional-level intelligence. By leveraging advanced technologies such as Large Language Models, reinforcement learning, and predictive analytics, the proposed system aims to create a proactive, scalable, and intelligent educational ecosystem.

The primary objective of this work is to design and evaluate a holistic agent-based framework that enhances learning outcomes, reduces educator workload, and supports strategic institutional decision-making. Through this approach, the study contributes to advancing the role of agentic AI in education and demonstrates its potential to redefine modern learning environments.

\section{Literature Review}

The application of Artificial Intelligence in education has evolved significantly over the past decade, transitioning from rule-based systems to more sophisticated machine learning and deep learning approaches. Early implementations primarily focused on intelligent tutoring systems and automated grading, which aimed to improve efficiency and provide limited personalization. While these systems achieved notable success in structured environments, they were constrained by their inability to adapt dynamically to changing user needs and contextual variations.

Recent research has introduced the concept of agentic AI, which extends traditional AI capabilities by incorporating autonomy, planning, and decision-making features. Kostopoulos et al.~\cite{b11} explored the role of agentic systems in education and highlighted their potential to deliver personalized learning experiences through adaptive tutoring and intelligent feedback. Similarly, Kamalov et al.~\cite{b12} examined agentic workflows involving reflection, planning, and tool utilization, demonstrating improved consistency compared to static AI models.

The increasing interest in agentic AI is further supported by comprehensive surveys such as Bandi et al.~\cite{b13}, which analyzed numerous studies and identified key characteristics including goal-driven reasoning, memory augmentation, and multi-agent collaboration. Acharya et al.~\cite{b14} extended this discussion by exploring applications of agentic systems across domains such as healthcare and finance, emphasizing adaptability and autonomous decision-making capabilities.

Despite these advancements, several challenges remain. Adabara et al.~\cite{b15} highlighted issues related to trust, governance, and security, emphasizing the need for robust frameworks to ensure ethical and reliable deployment. Similarly, Murugesan et al.~\cite{b16} discussed broader concerns such as transparency, accountability, and regulatory compliance, which are particularly critical in educational settings.

Further studies have explored the operational capabilities of agentic systems. Raheem et al.~\cite{b17} analyzed scalability and efficiency, noting improvements in decision-making processes. Pati et al.~\cite{b18} investigated goal-directed adaptive behavior, while Allam et al.~\cite{b19} demonstrated the use of agentic AI for real-time analytics and strategic planning in complex environments.

Additionally, Sapkota et al.~\cite{b110} provided a conceptual distinction between traditional AI agents and agentic AI systems, emphasizing the importance of persistent memory, coordination, and task decomposition in multi-agent environments. However, their study also identified limitations such as system brittleness, hallucination issues, and coordination failures.

Although existing literature demonstrates the transformative potential of agentic AI, most studies focus on isolated functionalities or single-layer implementations. There remains a significant research gap in developing unified frameworks that integrate student-level personalization, educator-level support, and institutional-level intelligence within a single cohesive system.

To address this gap, the present study proposes a comprehensive multi-agent architecture that unifies these components into a scalable and adaptive framework. By integrating reinforcement learning, predictive analytics, and large language models, the proposed system aims to overcome the limitations of existing approaches and provide a holistic educational support solution.

\section{Research Gap and Contributions}

Despite the rapid progress in Artificial Intelligence-driven educational technologies, existing systems remain largely limited in their ability to provide holistic and autonomous support across the educational ecosystem. Most current approaches focus on isolated functionalities such as intelligent tutoring, automated grading, or learning analytics, without establishing meaningful integration between students, educators, and institutional stakeholders. This fragmented design restricts the overall effectiveness of AI systems, as insights generated at one level are rarely utilized to inform decision-making at another.

Furthermore, traditional AI models in education predominantly operate in a reactive manner, relying heavily on user inputs and predefined workflows. Such systems lack the capability to proactively monitor learning behavior, anticipate academic risks, or adapt dynamically to evolving educational contexts. Although recent studies on agentic AI have introduced concepts such as autonomy, planning, and multi-agent collaboration, their application within education remains underexplored and often confined to single-agent or domain-specific implementations.

Another critical limitation lies in the absence of unified frameworks that combine advanced AI techniques, including reinforcement learning, predictive analytics, and large language models, within a coordinated multi-agent architecture. Existing works seldom address the challenge of seamless inter-agent communication, continuous feedback integration, and system-wide optimization. Additionally, issues related to scalability, interpretability, and real-time decision-making are not adequately addressed in current solutions.

These gaps highlight the need for a comprehensive, multi-level, and agent-driven framework capable of delivering proactive, adaptive, and scalable support across all components of the educational ecosystem.

To address the identified research gaps, this paper proposes a novel AUSS, designed as a multi-agent framework that integrates intelligence across student, educator, and institutional levels. Unlike existing approaches, the proposed system introduces a cohesive architecture that enables continuous interaction and coordination among autonomous agents, thereby facilitating a unified and intelligent educational environment.

The study presents a structured agent design based on four core functional modules, namely perception, reasoning, action, and evaluation, which collectively enable autonomous decision-making and adaptive learning. By incorporating large language models for content generation, reinforcement learning for policy optimization, and predictive analytics for performance assessment, the framework ensures both accuracy and adaptability in dynamic learning scenarios.

In addition, the proposed system introduces an event-driven communication mechanism that allows agents to share insights and trigger actions in real time, thereby improving responsiveness and coordination across different system layers. The framework is further validated through experimental analysis, demonstrating improvements in recommendation accuracy, grading efficiency, and dropout risk prediction, while maintaining low response latency and high scalability.

Overall, this work contributes to the advancement of agentic AI in education by providing a unified, scalable, and intelligent framework that bridges the gap between isolated AI applications and fully integrated educational ecosystems.

\section{Proposed Method}

This study proposes a multi-agent framework designed to deliver proactive, adaptive, and scalable support across the educational ecosystem. The proposed method integrates student-level personalization, educator-level automation, and institutional-level intelligence within a unified architecture, enabling continuous monitoring, intelligent decision-making, and dynamic adaptation.

\subsection{System Overview}

The AUSS framework operates through a coordinated network of autonomous agents, each responsible for specific functional objectives while maintaining continuous interaction with other agents. The system follows a layered architecture consisting of data acquisition, processing, agent intelligence, analytics, and application layers. Data collected from learning management systems, assessment records, and attendance logs are processed and transformed into meaningful features that drive agent decision-making.

The core of the system lies in its agent intelligence layer, where multiple agents collaboratively analyze data, generate insights, and execute actions. The system is designed to ensure seamless communication between agents through an event-driven mechanism, allowing real-time information exchange and coordinated responses to dynamic educational scenarios.

\subsection{Multi-Agent Architecture}

The proposed framework comprises three primary agents, each addressing a distinct level of the educational ecosystem. The Student Agent focuses on individual learners by analyzing performance, engagement, and behavioral patterns to generate personalized learning pathways and recommendations. It continuously monitors student progress and provides proactive feedback, enabling early identification of learning gaps.

The Educator Agent assists instructors by automating administrative tasks such as grading, report generation, and attendance tracking. It also provides analytical insights into class performance, helping educators identify at-risk students and adapt teaching strategies accordingly. Additionally, it supports content creation by generating quizzes, assignments, and instructional materials aligned with curriculum objectives.

The Institution Agent operates at a higher level, analyzing aggregated data to support strategic decision-making. It identifies trends related to student performance, resource utilization, and potential dropout risks, enabling institutions to optimize policies and improve operational efficiency. This agent also ensures compliance with institutional guidelines and ethical standards.

\subsection{Agent Functional Model}

Each agent within the AUSS framework follows a unified functional model consisting of four interconnected modules: perception, reasoning, action, and evaluation. The perception module is responsible for collecting and preprocessing data from various sources, transforming raw inputs into structured representations suitable for analysis. The reasoning module integrates multiple AI techniques, including large language models, rule-based logic, and reinforcement learning, to interpret data and determine optimal actions.

The action module executes decisions generated by the reasoning module, such as delivering recommendations, generating content, or triggering alerts. The evaluation module continuously monitors the outcomes of these actions, assessing their effectiveness based on predefined performance metrics. Feedback from this module is used to refine the system, enabling continuous learning and improvement.

\subsection{Learning and Decision Mechanism}

The proposed system employs a hybrid learning approach that combines predictive analytics, recommendation systems, and reinforcement learning. Collaborative filtering techniques are used to generate personalized learning recommendations by identifying similarities among learners. For performance prediction and engagement analysis, the system integrates machine learning models such as Random Forest for static features and Long Short-Term Memory (LSTM) networks for temporal patterns.

Reinforcement learning is used to optimize agent decision-making by learning policies that maximize long-term educational outcomes. The agent observes the current state of the system, selects an action, and receives feedback in the form of rewards, enabling it to improve its strategy over time. The policy update mechanism can be expressed as:

\begin{equation}
Q(s,a) \leftarrow Q(s,a) + \alpha \left[ r + \gamma \max_{a'} Q(s',a') - Q(s,a) \right]
\end{equation}
where $Q(s,a)$ represents the action-value function, $\alpha$ is the learning rate, $\gamma$ is the discount factor, $r$ is the reward, and $s'$ is the next state.

\subsection{Inter-Agent Communication}

To ensure coordination and scalability, the AUSS framework employs an event-driven communication mechanism that enables agents to exchange information in real time. When a significant event occurs, such as a decline in student performance or increased disengagement, the corresponding agent generates a trigger that is communicated to other agents. This allows the system to respond proactively, ensuring that interventions are timely and context-aware.

The communication mechanism also supports synchronization across agents, enabling consistent decision-making and preventing conflicts. This design enhances system robustness and ensures efficient handling of large-scale educational data.

\subsection{System Workflow}

\begin{figure}[h]
\centering
\begin{tikzpicture}[
node distance=2cm,
every node/.style={draw, rectangle, rounded corners, align=center},
arrow/.style={->, thick}
]

\node (data) {Data Sources\\(LMS, Attendance, Scores)};
\node (process) [below of=data] {Data Processing Layer};

\node (student) [below left=of process, xshift=-.5cm] {Student Agent};
\node (educator) [below of=process] {Educator Agent};
\node (institution) [below right=of process, xshift=1.2cm] {Institution Agent};

\node (analytics) [below of=educator] {Analytics \& Decision Layer};
\node (app) [below of=analytics] {Application Layer};

\draw[arrow] (data) -- (process);
\draw[arrow] (process) -- (student);
\draw[arrow] (process) -- (educator);
\draw[arrow] (process) -- (institution);

\draw[arrow] (student) -- (analytics);
\draw[arrow] (educator) -- (analytics);
\draw[arrow] (institution) -- (analytics);

\draw[arrow] (analytics) -- (app);

\draw[arrow, dashed] (student) -- (educator);
\draw[arrow, dashed] (educator) -- (institution);
\draw[arrow, dashed] (student) -- (institution);

\end{tikzpicture}
\caption{Architecture of the Proposed AUSS Multi-Agent System}
\end{figure}
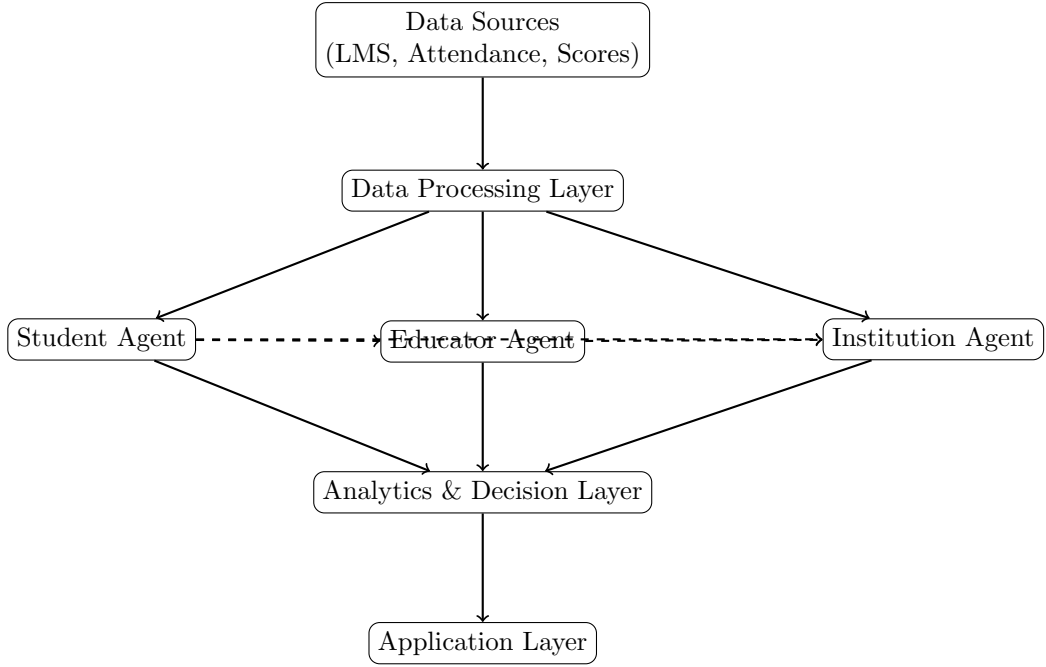

The overall workflow of the proposed system begins with data collection from various educational sources. The collected data are processed and forwarded to the respective agents, where they are analyzed using the perception and reasoning modules. Based on the analysis, agents generate actions such as recommendations, alerts, and reports, which are then evaluated for effectiveness. The feedback loop ensures continuous refinement of the system, enabling it to adapt to evolving learning environments.

Through this integrated approach, the proposed method provides a comprehensive solution that addresses the limitations of existing systems by combining autonomy, adaptability, and scalability within a unified agentic AI framework.

\begin{algorithm}
\caption{AUSS Multi-Agent Coordination Algorithm}
\begin{algorithmic}[1]

\STATE Initialize Student Agent $S_a$, Educator Agent $E_a$, Institution Agent $I_a$
\STATE Initialize environment state $s$
\STATE Initialize Q-values $Q(s,a)$ arbitrarily

\WHILE{system is active}

    \STATE Collect data from LMS, assessments, and attendance
    
    \STATE $S_a$ processes student data and predicts engagement/performance
    \STATE Generate personalized recommendations $R_s$
    
    \STATE $E_a$ analyzes class-level data
    \STATE Automate grading and generate reports $R_e$
    
    \STATE $I_a$ aggregates institutional data
    \STATE Detect risks and optimize resources $R_i$
    
    \STATE Identify system state $s$
    \STATE Select action $a$ using policy $\pi(s)$
    
    \STATE Execute action (recommendation / alert / automation)
    
    \STATE Observe reward $r$ and next state $s'$
    
    \STATE Update Q-value:
   \STATE $Q(s,a) \leftarrow Q(s,a) + \alpha [r + \gamma \max_{a'} Q(s',a') - Q(s,a)]$
    
    \STATE $s \leftarrow s'$

\ENDWHILE

\end{algorithmic}
\end{algorithm}

\section{Results and Discussion}

This section presents a comprehensive evaluation of the proposed Agentic Unified Student Support System, focusing on its predictive performance, system efficiency, and scalability. The experimental results demonstrate the effectiveness of the multi-agent framework in delivering accurate, timely, and adaptive educational support.

\subsection{Performance Evaluation}

The performance of the proposed system was assessed using key metrics such as accuracy, match rate, and F1-score across different agents and tasks. Table~\ref{tab:performance_new} summarizes the evaluation results.

\begin{table}[h]
\centering
\caption{Performance Evaluation of AUSS Components}
\label{tab:performance_new}
\begin{tabular}{lccc}
\toprule
Component & Task & Metric & Score (\%) \\
\midrule
Student Agent & Recommendation & Top-1 Accuracy & 92.4 \\
Student Agent & Prediction & Accuracy & 88.7 \\
Educator Agent & Grading & Match Rate & 94.1 \\
Institution Agent & Risk Detection & F1-score & 89.5 \\
\bottomrule
\end{tabular}
\end{table}

The results indicate that the proposed framework achieves consistently high performance across all system components. The Educator Agent exhibits the highest accuracy with a grading match rate of 94.1\%, demonstrating its reliability in automating evaluation tasks. The Student Agent also performs effectively in both recommendation and prediction tasks, indicating strong personalization capabilities. Furthermore, the Institution Agent achieves a high F1-score of 89.5\%, confirming its ability to detect at-risk students with high precision and recall.

\subsection{Accuracy Analysis}

\begin{figure}[h]
\centering
\includegraphics[width=0.45\textwidth]{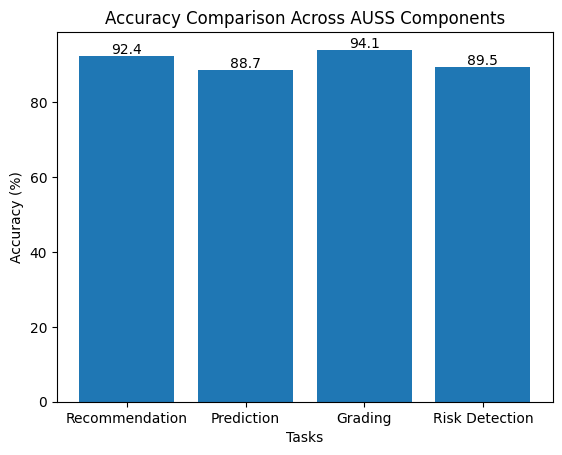}
\caption{Accuracy Comparison Across AUSS Components}
\label{fig:accuracy_new}
\end{figure}

Fig.~\ref{fig:accuracy_new} illustrates the comparative accuracy across different system components. It can be observed that the grading task achieves the highest accuracy, followed by recommendation and risk detection. The slightly lower prediction accuracy reflects the inherent complexity of forecasting student performance, which involves dynamic and temporal factors.

The overall accuracy trends demonstrate that the integration of collaborative filtering, machine learning models, and agent-based decision-making significantly enhances system performance.

\subsection{System Efficiency Analysis}

\begin{figure}[h]
\centering
\includegraphics[width=0.45\textwidth]{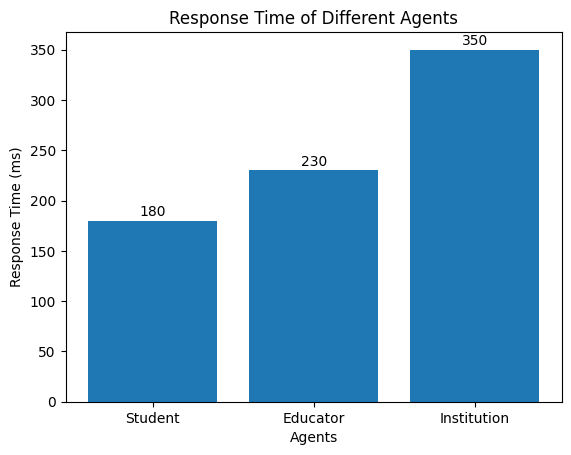}
\caption{Response Time of Different Agents}
\label{fig:response_new}
\end{figure}

The efficiency of the system was evaluated in terms of response time and computational load. Fig.~\ref{fig:response_new} presents the response time of different agents.

The Student Agent achieves the lowest response time of 180 ms, ensuring real-time interaction and feedback for learners. The Educator and Institution Agents exhibit slightly higher response times due to the complexity of their analytical tasks, yet remain within acceptable operational limits. These results confirm the system's capability to support real-time educational applications.

\subsection{Scalability and Load Distribution}

\begin{figure}[h]
\centering
\includegraphics[width=0.45\textwidth]{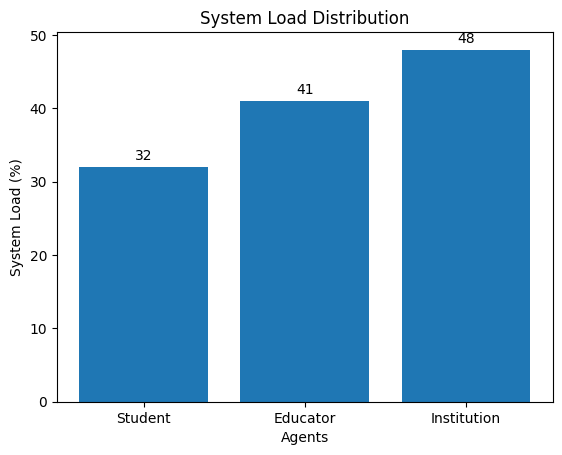}
\caption{System Load Distribution Across Agents}
\label{fig:load_new}
\end{figure}

Fig.~\ref{fig:load_new} illustrates the system load distribution across agents. The Institution Agent handles the highest load (48\%) due to its role in processing aggregated data and performing large-scale analytics.

Despite variations in workload, the system maintains a balanced distribution of computational resources, demonstrating its scalability and robustness in handling increasing data volumes.

\subsection{Discussion}

The experimental results clearly demonstrate the advantages of the proposed AUSS framework over traditional AI-based educational systems. The high accuracy across multiple tasks validates the effectiveness of the multi-agent architecture in delivering personalized and automated support. The integration of reinforcement learning enables adaptive decision-making, allowing the system to continuously improve based on user interactions.

Moreover, the low response times and efficient load distribution highlight the system's ability to operate in real-time environments without compromising performance. The collaborative interaction between agents plays a crucial role in enhancing system intelligence, as insights generated at one level are effectively utilized across other levels.

Overall, the proposed framework successfully addresses the limitations of existing systems by providing a unified, proactive, and scalable solution. The combination of accuracy, efficiency, and adaptability positions AUSS as a promising approach for next-generation intelligent educational systems.
\section{Conclusion and Future Work}

This paper presented the AUSS, a novel multi-agent framework designed to enhance educational processes through the integration of agentic Artificial Intelligence. By combining student-level personalization, educator-level automation, and institution-level intelligence within a unified architecture, the proposed system addresses the limitations of traditional reactive AI-based educational solutions. The framework leverages advanced techniques, including large language models, reinforcement learning, and predictive analytics, to enable proactive decision-making and continuous adaptation.

The experimental results demonstrate that the proposed system achieves high accuracy across multiple tasks, including personalized recommendation, automated grading, and dropout risk prediction. The Student Agent effectively delivers adaptive learning pathways, while the Educator Agent significantly reduces administrative workload without compromising evaluation quality. At the institutional level, the system provides valuable insights for strategic decision-making, particularly in identifying at-risk students and optimizing resource allocation. Furthermore, the system maintains low response times and balanced computational load, confirming its suitability for real-time and large-scale deployment.

Despite these promising outcomes, certain limitations remain. The current evaluation is conducted in a controlled experimental setting with limited-scale datasets, which may not fully capture the complexity of real-world educational environments. Additionally, while reinforcement learning enables adaptive decision-making, the design of reward functions and policy optimization strategies can significantly influence system behavior and requires further refinement. Issues related to data privacy, ethical AI deployment, and interpretability also warrant careful consideration, particularly in sensitive domains such as education.

Future work will focus on extending the proposed framework to real-world deployments across diverse educational institutions, incorporating large-scale and heterogeneous datasets to improve generalization. Advanced reinforcement learning techniques, such as multi-agent reinforcement learning and adaptive policy optimization, will be explored to further enhance system intelligence and coordination. Additionally, the integration of explainable AI mechanisms will be investigated to improve transparency and trust in automated decision-making. Strengthening privacy-preserving techniques, such as federated learning and secure data sharing, will also be considered to ensure compliance with ethical and regulatory standards.

Overall, this study establishes agentic AI as a transformative paradigm for intelligent education systems. The proposed AUSS framework provides a scalable, adaptive, and unified solution that has the potential to significantly improve learning outcomes, streamline educational processes, and support data-driven institutional governance in the evolving landscape of digital education.






\end{document}